\documentstyle[epsf]{aipproc}

\begin {document}

\title { Dynamical Behaviour in the Nonlinear Rheology
of Surfactant Solutions} 
\author{Ranjini Bandyopadhyay and A. K. Sood }
\address {Department of Physics,
Indian Institute of Science,
Bangalore 560 012,
India\\}
\maketitle
 
\begin{abstract}   
Several surfactant molecules self-assemble in solution to form 
long, flexible wormlike micelles which get entangled 
with each other, leading to viscoelastic gel phases.  
We discuss our recent work on the 
rheology of such a gel formed in the dilute aqueous 
solutions of a surfactant CTAT. 
In the linear rheology regime, the storage modulus $G^{\prime}(\omega)$ and 
loss modulus $G^{\prime\prime}(\omega)$ have been measured over a wide 
frequency range. In the nonlinear 
regime, the shear stress $\sigma$ shows a plateau as a 
function of the shear rate $\dot\gamma$ above a certain cutoff 
shear rate $\dot\gamma_c$.
Under controlled shear
rate conditions in the plateau regime, the shear stress
and the first normal stress difference show oscillatory time-dependence.
The analysis of the measured time series 
of shear stress and normal stress has been done using several 
methods incorporating state space reconstruction by embedding of time delay 
vectors.The analysis shows the existence of a finite correlation dimension 
and a 
positive Lyapunov exponent, unambiguously implying that the dynamics 
of the observed mechanical instability can 
be described by that of a dynamical system 
with a strange attractor of dimension varying from 2.4 to 2.9.
\end{abstract} 

\section{Introduction}

The amphiphilic nature 
of surfactant molecules results in  their reversible self-assembling
to form larger structures like spherical and 
cylindrical micelles, bilayers and vesicles. These supramolecular assemblies 
self-organise to form either long-range ordered liquid 
crystalline phases or short-range ordered isotropic liquid phases. Some 
examples of ordered phases are the nematic ordering of micellar discs or 
rods, 
the smectic A phase formed from the one-dimensional packing of infinite 
lamellae,
cubic phases formed by finite-size spherical micelles and 
two-dimensional hexagonal crystals formed by infinite cylindrical micelles.
In recent years, bicontinuous cubic phases have been observed in which a 
continuous bilayer or a monolayer forms a periodic, saddle-shaped, minimum
energy surface. In isotropic phases, microemulsions and sponge phases can 
form. Another example is the formation of viscoelastic gels from the
entanglement of very long and semiflexible cylindrical micelles with 
polymerlike behaviour. Interestingly, these micelles break and rejoin, 
giving rise to the name 'wormlike micelles'. The length distribution
depends on the surfactant and the salt concentrations, temperature and  
the energy of scission of the micelles. The objective of this paper is to
discuss our recent results on the nonlinear flow behaviour of such a system 
formed by the surfactant CTAT (cetyl trimethylammonium tosilate) in water.
At very low concentrations, ($c < 0.04$wt.\%) and above 
the Kraft temperature of 23$^{\circ}$C, CTAT molecules in 
water self-assemble to form spherical micelles exhibiting Newtonian flow. 
Above 0.9wt.\%, 
CTAT forms giant wormlike micelles. The hexagonal phase is seen at a
CTAT concentration of above 
27wt.\% \cite{sol1,sol2}. 

In polymeric melts, 
stress relaxation occurs primarily by reptation dynamics (curvilinear diffusion 
of the polymeric chain along the contour of an imaginary tube enclosing the chain), as suggested by de Gennes \cite{deg}, which defines a  relaxation 
time $\tau_{rep}$. In systems of giant wormlike micelles, there is an additional 
contribution to the stress relaxation from the reversible scission
(reversible breakage and recombination) of micelles, which occurs on a time
scale $\tau_b$. If $\tau_b << \tau_{rep}$, then  
stress relaxation in wormlike micelles  
occurs mainly by reversible scission and is described well 
by the Maxwell model which gives the response function 
G(t)=G$_{o}exp{(-\frac{t}{\tau_{M}})}$,  where G$_{o}$ is the high frequency
plateau modulus and $\tau_{M} = (\tau_{b}\tau_{rep})^{1/2}$ is the relaxation
time. The physical picture of the single relaxation time is a kind of 
motional narrowing \cite{can,cat}.

The reptation reaction model of wormlike micelles, 
which gives the Maxwell model for the linear viscoelasticity,
yields unusual nonlinear rheology observed in some 
viscoelastic gels (for example cetyl pyridinium chloride/
sodium salicylate) \cite{reh,gran}. The flow 
curve is characterised by a plateau 
in the shear 
stress versus shear rate curve\cite{reh}, while the normal stress is found to 
increase linearly with shear rate \cite{reh,spe}. These features in the flow
curve of giant wormlike micellar systems are associated with a flow 
instability of the shear banding type, where the sheared solution 
breaks up into bands of high and low viscosities supporting low and high 
shear
rates, respectively \cite{spe}. 
Flow birefringence \cite{mak} and nuclear magnetic resonance 
velocity imaging 
\cite{mai} have revealed the existence of banded flow 
in the shear stress plateau regime.
An alternate hypothesis explains the plateau in the shear stress as due to an isotropic-nematic phase transition 
that occurs  due to the application of high shear rates to the sample 
 \cite{ber}. 
Experiments on the transient rheology of CPyCl-NaCl after imposition of step
shear rates, show overshoots, oscillations and long time sigmoidal 
kinetics in the shear stress $\sigma (t)$ \cite{ber}. In our system of interest, 
CTAT 1.9wt.\% in water at 25$^{\circ}$C, we do not observe any overshoots in the 
stress relaxation on application of shear rates lying in the plateau 
region of the flow curve. However, we see that at high shear rates, the 
stress instead of decaying to a steady state, oscillates in time for a
long time. We have analysed the time series of the viscoelastic and normal 
stresses obtained from these experiments and have demonstrated the 
existence 
of deterministic, chaotic dynamics in sheared solutions of CTAT. 
Oscillations 
in the stress relaxation of CTAT have also been observed by Soltero et al at 
a concentration of 5wt.\% \cite{sol3}.  

Chaotic time series data have often been observed in experiments on 
fluid dynamics\cite{ott}. The Couette-Taylor flow in fluids and the 
Rayleigh-Benard convection in gases are two examples. 
Ananthakrishna et al \cite{ana} have shown the existence of chaotic dynamics in the 
jerky flow (Portevin-Le Chatelier effect) of some metal alloys undergoing 
plastic deformation. These systems also exhibit nonmonotonic flow 
curves.
Deterministic chaos represents the apparently irregular behaviour 
of dynamical systems that arises from strictly deterministic laws in the 
absence of any external stochasticity. Chaotic dynamics in 
physical systems is 
characterised by an exponentially sensitive dependence on initial conditions, 
as a result of which long-term predictability of the dynamics of these 
systems is impossible. Chaotic time series may be characterised by certain 
invariants, metric and dynamical, such as the various fractal 
dimensions, the largest
Lyapunov exponent and the Kolmogorov entropy. A positive value of the largest
Lyapunov exponent is a direct 
consequence of the sensitivity of the trajectories 
in phase space to small changes in 
the initial conditions. The fractal dimensions 
provide a measurement of the topology 
of the attractor on which all the trajectories asymptotically converge. Apart from describing the complexity of the attractor, they 
also provide a measure of the regularity of occurrence  of the phase points 
on its surface \cite{abar}. 

In this paper, we will show the existence of a positive Lyapunov exponent 
and
a finite correlation dimension in the observed time series of the 
stress relaxation of CTAT
on application of step shear rates lying in the plateau region of the 
flow curve. Further, the magnitudes of the invariants are found to 
increase monotonically 
with shear rate, the control parameter in our experiments. 

\section{Experimental Details} 

CTAT (purchased from Sigma Chemicals, India) 
samples were prepared by dissolving
1.9wt.\% of powdered CTAT in distilled and deionised water. 
The samples were kept in an incubator at a temperature of 60$^{\circ}$C 
for a week, and frequently shaken to facilitate homogenisation. 
The samples were then kept at the experimental temperature (25$^{\circ}$C)
for two days. The rheological properties of the CTAT thus prepared and 
equilibrated were measured using a Rheolyst AR-1000N ( T. A. Instruments, 
U. K.) stress-controlled rheometer with temperature control and software 
for shear rate control. The rheometer was also equipped  with four strain 
gauge transducers at a distance of $R$ from the centre 
of the plate for the measurement of the normal force  $F_{z}$.
$F_{z}$ is related to the first normal stress difference Z by Z = 
${\frac{2F_{z}}{\pi R^{2}}}$
\cite{mac}. All the measurements 
reported in this paper were performed 
using a cone-and-plate geometry \cite{mac} of 
diameter 4cm and cone angle 1$^{\circ}$59". In each flow 
experiment, the rheometer can collect upto a maximum of 1500 data points, 
with a time gap of 1s or more between acquisition
of successive data points. Sample history effects were 
found to be important in CTAT and hence all experiments have been 
done on fresh samples from the same batch. 
Most of the experiments were performed at 25$^{\circ}$C and a few at 
35$^{\circ}$C.

\section{Results}

Fig. 1 shows the frequency response measurements  
at  (a) 25$^{\circ}$C and  (b) 35$^{\circ}$C. The
linear regime was first ascertained by imposing stresses between
0.005 Pa to 2 Pa, oscillating at a frequency of 0.1 Hz, and finding the
window of the applied stress values over which  the elastic modulus
G$^{\prime}(\omega)$
and the
viscous modulus G$^{\prime\prime}(\omega)$ are constant. 
From this measurement, the oscillatory stress for the frequency response 
measurements at both the temperatures has been chosen to be 0.6 Pa. 
\begin{figure}
\centerline{\epsfxsize = 8cm \epsfysize = 8cm \epsfbox{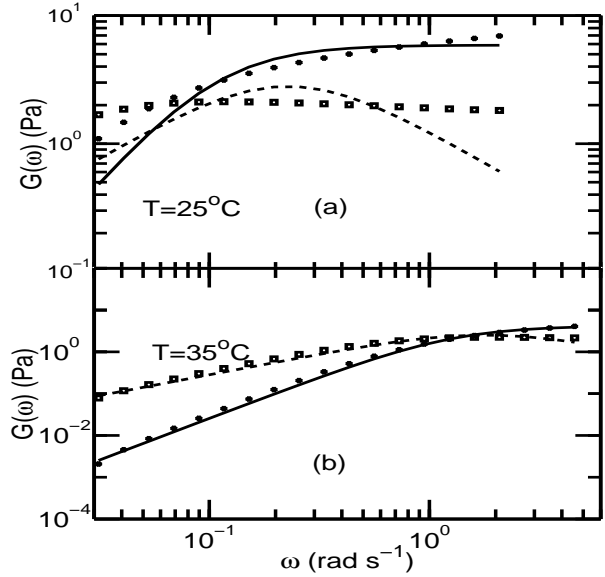}}
\caption{ Frequency response of CTAT 1.9wt.\% at (a) 25$^{\circ}$C and 
(b) 35$^{\circ}$C, at an oscillatory stress of 0.6Pa. G$^{\prime} (\omega)$ 
is indicated by solid circles and G$^{\prime\prime} (\omega)$ by hollow 
squares. The solid and dashed lines show the fits to the Maxwell 
model for G$^{\prime} (\omega)$ and G$^{\prime\prime} (\omega)$, respectively. }
\label{Fig.1}
\end{figure}
In 
Figs. 1(a) and 1(b), the solid circles and the hollow squares represent 
G$^{\prime}(\omega)$ and G$^{\prime\prime}(\omega)$, respectively. 
The dominant relaxation time,
estimated from the angular frequency at which G$^{\prime}(\omega)$
and G$^{\prime\prime}(\omega)$ are equal, is found to be 17s (1s) at 
25$^{\circ}$C (35$^{\circ}$C).  The 
solid (G$^{\prime}(\omega)$) and dashed lines (G$^{\prime\prime}(\omega)$)
are the corresponding fits  
to the Maxwell model, given by 
$G^{\prime}(\omega)=\frac{G_{o}(\omega \tau_{M})^{2}}{1+(\omega \tau_{M})^{2}}$
and  
$G^{\prime\prime}(\omega)=\frac{G_{o} \omega \tau_{M}}{1+(\omega \tau_{M})^{2}}$. 
As mentioned before, 
for viscoelastic gels of wormlike micelles formed by 
the CTAC-NaSal,CPyCl-NaSal and CPySal-NaSal systems  \cite{can,gran}, 
the Maxwell model works very well. 
However, for CTAT at 25$^{\circ}$C, 
the fit of the data to the Maxwell model
is rather poor. Interestingly, the fit is reasonably good 
for experiments done at 35$^{\circ}$C. This may be due to the fact
that the motional narrowing condition, $\tau_{b}<<\tau_{rep}$, is
not satisfied at low temperatures.

\begin{figure}
\centerline{\epsfxsize = 9cm \epsfysize = 9cm \epsfbox{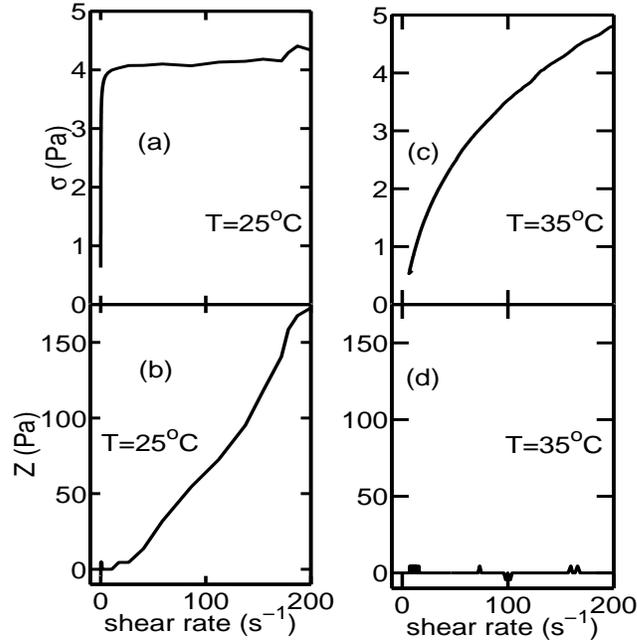}}
\caption{ The metastable flow curves of CTAT 1.9wt.\%, showing the 
shear stress and normal stresses, respectively, as a function of shear 
rate at 25$^{\circ}$C ((a) and (b)) and 35$^{\circ}$C ((c) 
and (d)).}
\label{Fig.2}
\end{figure}

\begin{figure}
\centerline{\epsfxsize = 15cm \epsfysize = 15cm \epsfbox{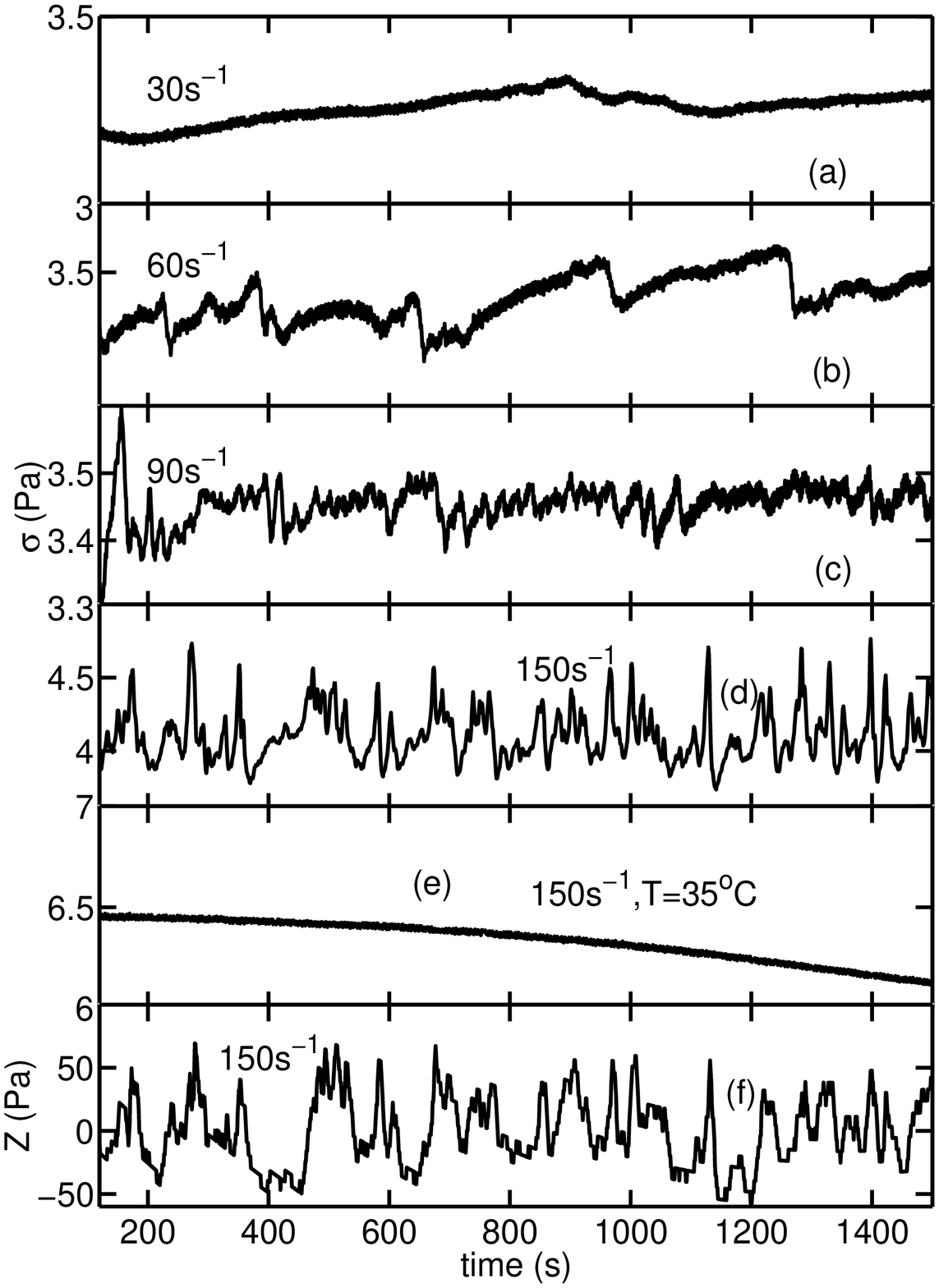}}
\caption{Shear stress relaxation measured in CTAT 1.9wt.\% 
 at 25$^{\circ}$C on application 
of a step shear rate of (a) 30s$^{-1}$, (b) 60s$^{-1}$, (c) 90s$^{-1}$ and 
(d) 150s$^{-1}$. (e) shows the relaxation of shear stress at 35$^{\circ}$C
 and $\dot\gamma$ = 150s$^{-1}$.(f) shows the relaxation of normal stress 
at 25$^{\circ}$C on application of a shear rate of 150s$^{-1}$.}
\label{Fig.3}
\end{figure}

Fig 2 shows the viscoelastic stress and normal stress measurements when 
the stress was increased from 10 mPa to 4.7 Pa 
after allowing a time $\tau_{M}$ between acquisition of consecutive 
data points, and measuring the corresponding shear rate. 
These measurements give the metastable branches of the flow curve 
at 25$^{\circ}$C (Fig. 2(a)) and 35$^{\circ}$C (Fig. 2(c)). 
Figs. 2(b) and 2(d) show the measured normal stresses as a function of 
increasing shear rate at both the temperatures. 
At 25$^{\circ}$C, the viscoelastic stress 
(2(a)) shows a crossover to a plateau, 
while the normal stress increases quadratically (2(b)) with increase 
in shear rate. 
The presence of a plateau in the viscoelastic stress at high shear rates 
is a signature of a mechanical
instability, possibly of the shear banding type \cite{gran,spe}. 
We have ruled out the possibility of an 
isotropic-nematic  phase transition on account of the very 
low volume fractions 
of CTAT used in our experiments. At 35$^{\circ}$C, the viscoelastic stress 
does not show a plateau region and the normal stress is roughly zero, 
which implies the absence of instability
in sheared solutions of wormlike micelles at higher 
temperatures \cite{por}. 

We have studied in great detail the relaxation of the viscoelastic stress 
on application of step shear rates lying in the plateau region of the 
flow curve. Fig. 3 (a)-(d) show the stress relaxation 
measured for 1500 seconds, after subjecting the sample to 
step shear rates of 30, 60, 90 and 150s$^{-1}$ (Figs. 3 (a)-(d)) at 
25$^{\circ}$C. 
The stress decays initially for t$\sim$100 seconds 
(not shown in Fig. 3), and then oscillates in time. 
This time-dependent behaviour of the stress becomes more pronounced at 
higher shear rates. Fig. 3(e) show the stress relaxation at 35$^{\circ}$C 
on application of a step shear rate $\dot\gamma$ $=$ 150s$^{-1}$. 
The oscillations seen at 25$^{\circ}$C at $\dot\gamma$=150s$^{-1}$ 
are found to disappear completely at 35$^{\circ}$C, 
leading us to ascertain that the time-dependent 
oscillations observed in the stress at 25$^{\circ}$C 
are a direct consequence of the mechanical instability in the system. 
The first normal stress difference Z, which is measured simultaneously
with the viscoelastic stress $\sigma$ 
at $\dot\gamma$=150s$^{-1}$ also shows oscillations, as shown in Fig.3(f).
In what follows, we describe how
we have analysed the time-dependent data in $\sigma$ and Z to establish the
presence of chaotic dynamics in the sheared CTAT solution. The time series
analysis has been performed by using the method of state space construction
by embedding time delay vectors \cite{ruel}. 
Suppose \{$\sigma_j=\sigma(j\Delta$ t), j=1,1500\} denotes the 
time series in stress $\sigma$ on application of a step shear rate. 
$\Delta t =$  1s is the 
time elapsed between collection of data points.
In this method, we construst m-dimensional L-delay vectors given by 
\{$\vec X_i=(\sigma_i, \sigma_{i+L}, \sigma_{i+2L} ,....,\sigma_{i+(m-1)L}$)\}.
\{$\vec X_i$:i=1,2,...,N-(m-1)L\} defines a trajectory in m-dimensional space 
due to the presence of a dynamics 
F$(\vec  X_i):\vec X_i \rightarrow \vec X_{i+1}$. 
In doing so, it is crucial to choose the optimal embedding 
dimension $m_{o}$ and the optimal delay time $L_{o}$ properly\cite{tak}. 
We have done this by analysing the local exponential divergence 
plots and the plots of 
$\gamma$ = $<ln(||\vec X_{i+k}-\vec X_{j+k}||/||\vec X_{i}-\vec X_{j}||)>$ 
versus L 
for the 
trajectories obtained by embedding 
the time series in m dimensions. Here $<..>$ 
denotes averaging over all $(i,j)$ pairs. Embedding the trajectories in 
higher dimensions leads to a reduction in the number of false 
neighbours as at m $< m_{o}$, we are  not looking at the real
dynamics, but at its projection. At m$> m_{o}$, the number of false 
neighbours is zero.
The left panel of Fig 4 shows the local exponential divergence 
plots for $\dot\gamma$=150s$^{-1}$, where the abscissa is ln($||\vec X_{i}-\vec X_{j}||$) and 
the ordinate is ln($||\vec X_{i+k}-\vec X_{j+k}||/||\vec X_{i}-\vec X_{j}||$)  
for (L,m), where L= 1 and 5 and m=3 to 6.
$||\vec X_{i}-\vec X_{j}||$ is the distance between the 
$i$th and $j$ th trajectories in m-dimensional space and
is chosen to be smaller than a prescribed small distance r$^{\star}$.
$||\vec X_{i+k}-\vec X_{j+k}||$ is the Euclidian distance  between 
$\vec X_{i}$ and $\vec X_{j}$ after k iterations of the dynamics F. 
For our  analyses, we have chosen $r^{\star}$ to be equal to
4\% of the maximum value of $||\vec X_{i}-\vec X_{j}||$. 
In all our calculations,
k has been 
determined from the value of t at which the autocorrelation function 
$C_{L}(t)$ of the time series in $\sigma$ or Z 
falls to 1/e of its peak value. 
The plots are found to get increasingly compact at L=5 
as m increases and 
there is no significant change as we go from m=5 to m=6. Hence we conclude 
that the optimal embedding dimension $m_{\circ}$ for our time series in
$\sigma (t)$ at $\dot\gamma =$ 150s$^{-1}$ is 5.
 
\begin{figure}
\centerline{\epsfxsize = 15cm \epsfbox{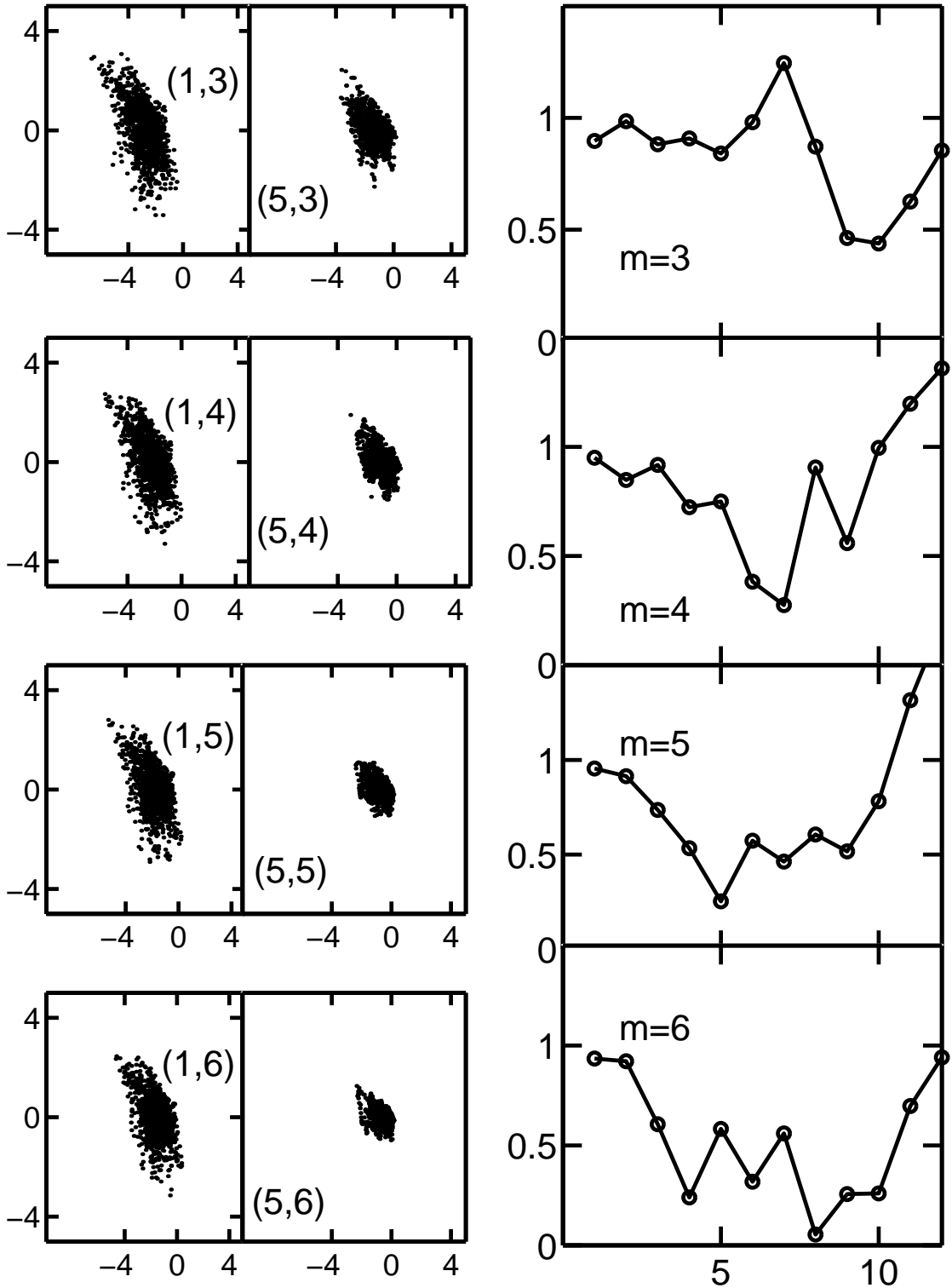}}
\caption{The left panel shows the local exponential divergence plots for 
typical values of m and L, marked as (L,m) for $\dot\gamma$=150s$^{-1}$. The abscissa is 
ln($||X_{i}-X_{j}||$) 
and the ordinate is ln($||X_{i+k}-X_{j+k}||/||X_{i}-X_{j}||$). The 
right panel 
shows the plots of $\gamma$ versus L at m=3 to 6, also for 
$\dot\gamma$=150s$^{-1}$. $\gamma$ has a 
minimum at m=5, L=5.}
\label{Fig.4}
\end{figure}

After obtaining a value of m from the local exponential divergence
plots, we have plotted 
$\gamma$ versus L (right panel, Fig. 4) for m = 3 to 6. As m increases, 
$\gamma$ is found to decrease, and there is no 
significant change in the $\gamma$ versus L plot
on changing m from m=$m_{\circ}$ (=5) to $m_{\circ+1}$ (=6). We see that $\gamma$ 
has a minimum
at L = 5 for m = 5 and hence we have chosen the optimal delay time
$L_{o}$ to be 5 and the optimal embedding dimension $m_{o}$ to be 5 for 
$\dot\gamma$=150s$^{-1}$. 

\begin{figure}
\centerline{\epsfxsize =10cm \epsfysize = 10cm \epsfbox{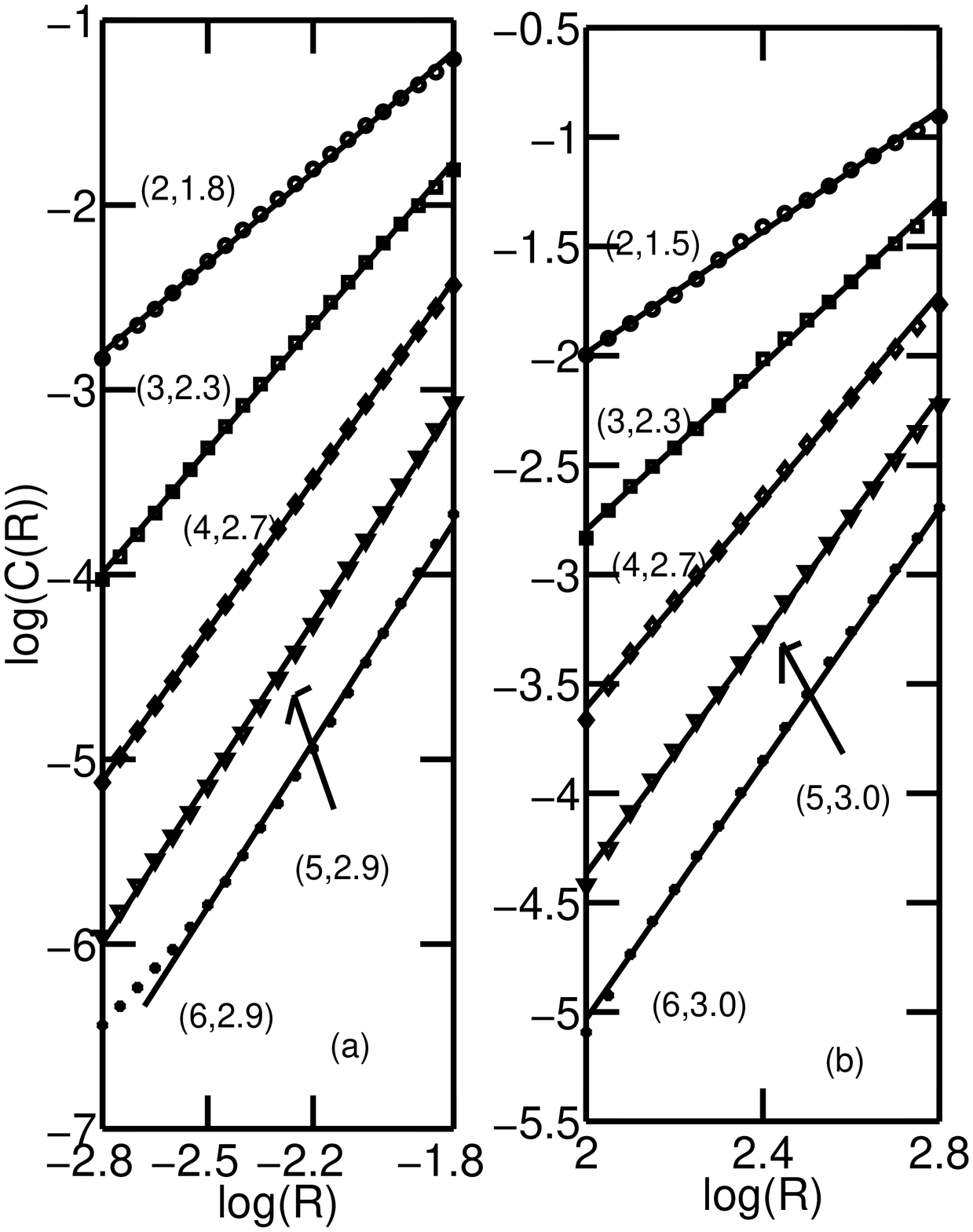}}
\caption{ Calculation of the correlation dimensions $\nu$ from the 
time series corresponding to 
(a) shear stress $\sigma$ and (b) normal stress 
Z at 25$^{\circ}$C and $\dot\gamma$=150s$^{-1}$, for m=2 to 6 
(denoted by (m,$\nu$)).
The values of $\nu$ calculated from 
the slopes of the plots of log(C(R)) versus log(R) are 2.9 for $\sigma$ and 3.0 
for Z.}
\label{Fig.5}
\end{figure}

\begin{figure}
\centerline{\epsfxsize =8cm \epsfysize = 8cm \epsfbox{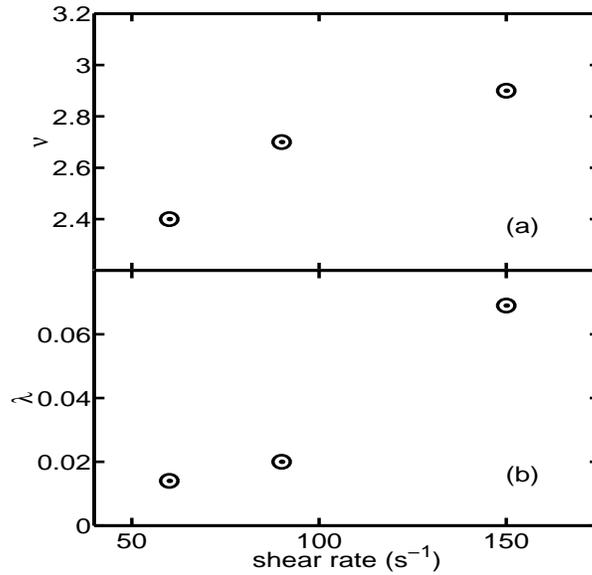}}
\caption{(a) The correlation dimension $\nu$ and (b) the 
Lyapunov exponent $\lambda$ as a function of the shear rate.}
\label{Fig.5}
\end{figure}

\begin{figure}
\centerline{\epsfxsize =8cm \epsfysize = 8cm \epsfbox{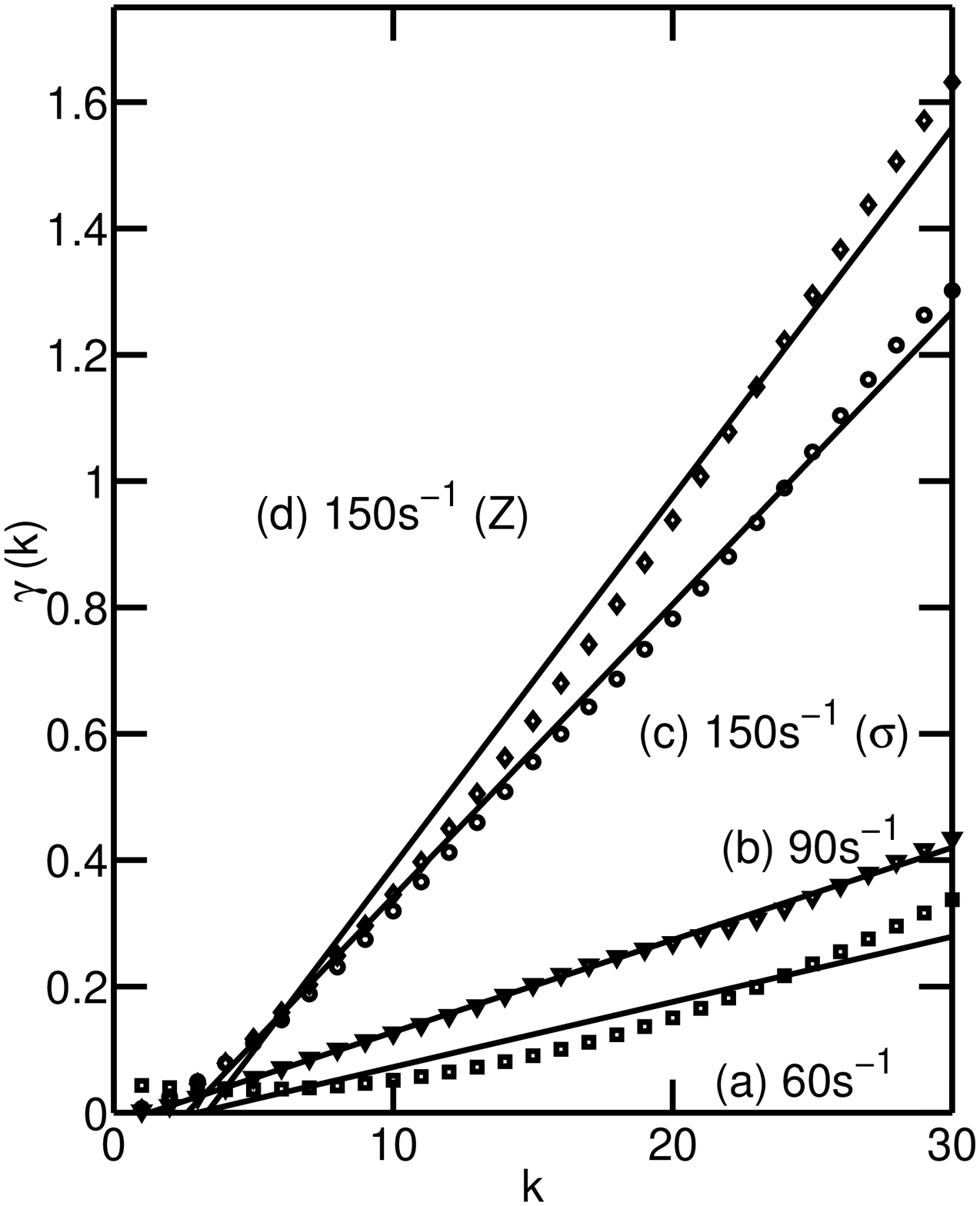}}
\caption{ Plot of $\gamma$ versus k for the time series of the shear 
stresses $\sigma$ at (a) $\dot\gamma$=60 (m=5,L=11), (b)90 (m=5,L=9) and 
(c)150s$^{-1}$(m=5,L=5). 
(d) shows the same for the time series of the normal stress Z 
at $\dot\gamma$=150s$^{-1}$ (m=5,L=8). The Lyapunov 
exponents which can be estimated from the slope 
of the curves are found to be positive for all shear rates shown. }
\label{Fig.6}
\end{figure}

We have calculated the correlation dimensions $\nu$ of the attractors 
in state space on which the trajectories asymptotically lie by 
computing the correlation integral in m-dimensional space  
$C(R)={lim\atop N\rightarrow \infty }{1\over N^{2}}\sum
_{i,j=1}^{N}H({R-||\vec X}_{i}-\vec X_{j}||)$ \cite{gra} for m=2 to 6. In the 
expression for C(R),  H(x) is the Heaviside function and 
$||\vec X_i-\vec X_j||$ is the 
distance between the pair of points ($i$,$j$) in the m-dimensional
embedding space. The sum in the above expression
gives 
the number of pairs of trajectories separated by a distance less than R. 
For small 
R's, C(R) is known to scale as 
$C(R)\sim R^{\nu }$,                           
where  the correlation dimension $\nu$ gives  useful 
information about the local structure of the attractor \cite{gra}. The plot 
of C(R) versus R has a constant slope over a given range of R, 
which does not change 
on increasing m from m$_{\circ}$ to m$_{\circ +1}$ and 
is denoted by $\nu$. If $\nu<$ m, then the signal exhibits deterministic 
chaos. Fig. 5 shows the calculation of the correlation dimensions for the 
time series in viscoelastic stress $\sigma$ and normal stress Z at
$\dot\gamma$=150s$^{-1}$. 
The slope  of the C(R) versus R plot in the plateau region is found to 
saturate at m=5 in both  cases yielding correlation dimensions $\nu$ of 
2.9 and 3, respectively (Figs. 5(a)-(b)). 
The satisfaction of the condition $\nu<$m, and the value of 
$\nu>$2 points to the existence of deterministic chaotic dynamics 
\cite{gra} in the stress relaxation of CTAT. We have also calculated the 
correlation dimensions of the attractors corresponding to the dynamics of 
stress relaxation at 60s$^{-1}$ and 90s$^{-1}$. The calculated values of 
$\nu$ in the two cases are 2.4 and 2.7 respectively, indicating the 
presence of chaotic dynamics at these shear rates. Fig 6(a) shows the 
calculated values of correlation dimensions $\nu$ as a function of the 
shear rate. $\nu$ is found to increase monotonically with the control 
parameter (shear rate in our case), similar to that 
observed in the  weakly turbulent Couette-Taylor flow 
exhibited by orange oil\cite{swin}, where the Rayleigh number 
was the control parameter.

A chaotic time series is characterised by a positive Lyapunov exponent, 
which describes the divergence of neighbouring trajectories in state space 
\cite{ott}. We have calculated the largest 
Lyapunov exponents for the time 
series of the viscoelastic stresses at 60s$^{-1}$, 90s$^{-1}$ and 
150s$^{-1}$ and the normal stress at 150s$^{-1}$ (Fig 7) using the method 
proposed by Gao and Zheng \cite{gao}. This method requires the calculation 
of $\gamma (k)$ as a function of k, with 
$||\vec X_{i}-\vec X_{j}|| \le r^{\star}$ and $\vec X_{i}$ and $\vec X_{j}$ 
chosen such that $|j-i| \ge \omega$, where $\omega = (m-1)L$,
 to exclude 
tangential motion.
A linear fit of $\gamma (k)$ versus k yields a positive slope S and 
nearly zero intercept, from which the Lyapunov exponent $\lambda$ has 
been extracted by using $\lambda =  S/\Delta t ln2$. $\lambda$ is found 
to increase monotonically with shear rate as shown in Figure 6(b).

\section{Conclusions}

The existence of a finite correlation dimension $\nu$ and a 
positive Lyapunov exponent $\lambda$ indicates the existence of 
deterministic chaos in the dynamics 
of stress relaxation in CTAT . This occurs only when the shear rates 
are high enough and lie in the plateau region of the
flow curve. Since the volume fraction of CTAT in our
experiment is very small, we rule out the possibility 
of an isotropic-nematic 
phase transition as the cause of the observed instabilities.
We conclude that the chaotic dynamics 
is a natural consequence of the mechanical instability.
The calculations described above predict an optimal 
embedding dimension $m_{o}$ 
of 5 for 
the description of the dynamics of the system. We have previously
established the presence of chaotic dynamics in 
the stress relaxation in CTAT 1.35wt.\%, when the sample was subjected
to step shear rates lying in the plateau region of the flow curve\cite{ran}.
However, at concentrations higher than 2.5wt.\%, it is difficult to obtain 
a sufficiently long time series, as the large normal stresses that develop 
in the sample \cite{sol2,spe} cause its
expulsion from the rheometer.
 
 In order to theoretically model the dynamical aspects of the observed
phenomenon in viscoelastic gels, we need to set up 
space and time-dependent, coupled, nonlinear differential equations in 
at least 5 dynamical variables. One possible choice could be the 
Johnson-Segalman (J-S) equation 
\cite{joh} which is the solution of the Navier-Stokes
equation coupled to the continuity equation for
isothermal, incompressible and viscoelastic fluids. 
 The viscoelastic nature
of the polymer is accounted for by writing the total stress as the sum of 
a Newtonian part and a deformation history-dependent viscoelastic part. 
This model yields a non monotonic flow curve and oscillations in the 
start-up stress in polymeric systems\cite{mal}. This model, however,
has not been analysed to predict
the chaotic time-dependence of the stresses that is seen 
in our experiments. 
We believe that the flow-concentration coupling and the dynamics of the 
mechanical interfaces between 
shear bands need to be incorporated in the J-S model. 
 Further, the reversible breakdown and
recombination of micelles also need to be considered. A model constructed
by incorporating these additional features, which takes into account the 
nonlinear coupling between the relevant dynamical variables 
like shear and normal stress, shear rate and concentration profiles 
 is likely to exhibit 
the chaotic behaviour that we observe.  

\section{Acknowledgements}
The authors thank Dr. Geetha Basappa 
for her participation in the initial part 
of the project and Dr. V. Kumaran, Dr. P. R. Nott and Dr. S. Ramaswamy for the 
use of the rheometer. AKS thanks the Board of Research in Nuclear Sciences
and RB thanks CSIR, India for financial support

\end{document}